# Magnetic kagome materials RETi$_3$Bi$_4$ family with weak interlayer interactions

Jingwen Guo[1,2,#], Liqin Zhou[3,4,#], Jianyang Ding[5,#], Gexing Qu[3,4], Zhengtai Liu[6], Yu Du[1,2], Heng Zhang[1,2], Jiajun Li[1,2], Yiying Zhang[1,2], Fuwei Zhou[1,2], Wuyi Qi[1,2], Fengyi Guo[1,2], Tianqi Wang[1,2], Fucong Fei[1,7,8]*, Yaobo Huang[6], Tian Qian[3], Dawei Shen[9], Hongming Weng[3,4,10]*, Fengqi Song[1,8]*

[1] *National Laboratory of Solid State Microstructures, Collaborative Innovation Center of Advanced Microstructures, Nanjing University, Nanjing 210093, China.*

[2] *School of Physics, Nanjing University, Nanjing 210093, China.*

[3] *Beijing National Laboratory for Condensed Matter Physics, Institute of Physics, Chinese, Academy of Sciences, Beijing 100190, China.*

[4] *University of Chinese Academy of Sciences, Beijing 100049, China.*

[5] *State Key Laboratory of Functional Materials for Informatics, Shanghai Institute of Microsystem and Information Technology (SIMIT), Chinese Academy of Sciences, Shanghai 200050, China.*

[6] *Shanghai Synchrotron Radiation Facility, Shanghai Advanced Research Institute, Chinese Academy of Sciences, Shanghai 201210, China.*

[7] *School of Materials Science and Intelligent Engineering, Nanjing University, Suzhou 215163, China.*

[8] *Atom Manufacturing Institute, Nanjing 211806, China.*

[9] *National Synchrotron Radiation Laboratory, University of Science and Technology of China, 230029 Hefei, China.*

[10] *Songshan Lake Materials Laboratory, Dongguan, Guangdong 523808, China.*

[#] These authors contributed equally to this work.
* Corresponding authors. E-mail: feifucong@nju.edu.cn; hmweng@iphy.ac.cn; songfengqi@nju.edu.cn;

**Abstract**

Kagome materials have attracted a surge of research interest recently, especially for the ones combining with magnetism, and the ones with weak interlayer interactions which can fabricate thin devices. However, kagome materials combining both characters of magnetism and weak interlayer interactions are rare. Here we investigate a new family of titanium based kagome materials $RETi_3Bi_4$ (RE = Eu, Gd and Sm). The flakes of nanometer thickness of $RETi_3Bi_4$ can be obtained by exfoliation due to the weak interlayer interactions. According to magnetic measurements, out-of-plane ferromagnetism, out-of-plane anti-ferromagnetism, and in-plane ferromagnetism are formed for RE = Eu, Gd, and Sm respectively. The magnetic orders are simple and the saturation magnetizations can be relatively large since the rare earth elements solely provide the magnetic moments. Further by angle-resolved photoemission spectroscopy (ARPES) and first-principles calculations, the electronic structures of $RETi_3Bi_4$ are investigated. The ARPES results are consistent with the calculations, indicating the bands characteristic with kagome sublattice in $RETi_3Bi_4$. We expect these materials to be promising candidates for observation of the exotic magnetic topological phases and the related topological quantum transport studies.

**Introduction**

Materials with kagome lattices have been studied for a long time in the condensed matter physics[1-4]. Kagome lattice is made of two-dimensional honeycomb network of corner-sharing triangles[5]. Based on the simple tight-binding model[6], it can host flat bands, Dirac cones, as well as saddle points, which have been theoretically predicted and experimentally observed [7-10]. Combining with spin-orbit coupling (SOC) and magnetism, plenty of topological phases have been realized in kagome materials and fruitful novel properties have emerged. For example, large intrinsic anomalous Hall effect (AHE) arising from Berry curvature has been observed in Weyl semimetal $Co_3Sn_2S_2$ and nonlinear antiferromagnets $Mn_3X$ (X = Sn, Ge)[11-15]. Flat band and Dirac cones have been discovered in $CoSn$[10,16] and $Fe_xSn_y$ family[9,17,18]. In recent years, kagome magnet system $RT_6X_6$ (R = rare earth; T = V, Cr, Mn, Fe, Co; X = Ge, Sn) with crystallizing in the $MgFe_6Ge_6$ structure has also attracted immense interest due to the hosted massive Dirac fermions, flat band along with Dirac cones, and the large intrinsic anomalous Hall effect[19-26]. Although the material candidates with kagome lattices varies, most of their interlayer interactions are strong, which makes the synthesized crystals bulky, and the properties of the kagome layer inside these materials deviate from the raw two-dimensional kagome lattices. The bulky 3D crystals are difficult to fabricate thin film devices on certain substrates and to tune the physical properties of the materials by gate voltage. Therefore, people are focusing on the discovery of kagome materials with weak interlayer interactions. In 2019, a new family materials $KV_3Sb_5$, $RbV_3Sb_5$, and $CsV_3Sb_5$ were discovered[27]. In this family, considering the weak chemical bonds between V, $Sb_2$, and alkali metal atoms, the crystals are cleavable by knives or scotch tapes[27], and thin flakes can be exfoliated on substrates such as silicon wafers and $Al_2O_3$, and mesoscopic transport devices can be fabricated[28-30]. Superconductivity and charge density wave have also been found in this family of materials[27,31-37], which has caused great interest of researchers in condensed matter physics and material science. However, comparing with the multiple types of magnetism and the fruitful researches on magnetic topological phases and anomalous Hall effect in kagome materials with strong interlayer interactions such as $Fe_3Sn_2$, $Mn_3Sn$, $FeSn$, $Co_3Sn_2S_2$, and $TbMn_6Sn_6$ family, and there is no spontaneous magnetization among these kagome materials of $KV_3Sb_5$ family with weak interlayer interactions. Therefore, finding a kagome material with weak interlayer interactions which combines magnetism, kagome lattices, cleavable crystals, and thin film device availability, would greatly benefits the exploration of more kagome physical properties and fruitful magnetic topological phases.

In this work, we investigate a newly discovered kagome materials $RETi_3Bi_4$ (RE = Eu, Gd, and Sm). In this family, titanium atoms form the kagome lattice. The interlayer interactions between the adjacent RE/Bi layers are weak due to the limited spatial density of the RE-RE zig-zag chain shaped chemical bonds. By mechanical exfoliation by tapes, the crystals of all three materials can be exfoliated into thin films with thickness as low as nanometer scale. The magnetic properties of $RETi_3Bi_4$ family are also systematically measured. Unlike the previous studied Mn or Fe based kagome lattices, the spontaneous magnetization is not provided by kagome lattice itself formed by transition metals, but solely by rare earth elements. Simple magnetism of out-of-plane ferromagnetism (FM), out-of-plane anti-ferromagnetism (AFM), and in-plane ferromagnetism are formed for RE = Eu, Gd, and Sm respectively. Combining with the first-principle calculations and angle-resolved photoemission spectroscopy (ARPES), the band structures of this family of compounds are also investigated. We take $EuTi_3Bi_4$ as the representative to elucidate the exotic electronic properties of these materials with kagome sublattices. We observed the distinct bands with Dirac cone-like dispersions and the flat bands above the Fermi level by calculation, and the ARPES results are consistent with the calculations for the band dispersions below Fermi energy. $RETi_3Bi_4$ family of kagome compounds combines cleavable lattice structure and tunable but simple magnetisms. This family provides a promising material platform for promoting the further explorations on the exotic magnetic topological phases and the related topological quantum transport studies.

**Results**

The structure of $RETi_3Bi_4$ is displayed in Fig. 1(a), and Fig. 1(b) shows the clean kagome layer of Ti with distorted slightly from the identical positions, less than 0.1 Å for $EuTi_3Bi_4$ for example. $RETi_3Bi_4$ holds larger prototype structures which have four kagome layers per unit cell, compared with other kagome materials like $RT_6X_6$ (R = rare earth; T = V, Cr, Mn, Fe, Co; X = Ge, Sn) and $AV_3Sb_5$ (A = K, Rb, Cs), and forms a orthogonal lattice structure with *Fmmm* space group (No. 69)[38-40]. The interlayer interactions between the adjacent RE/Bi layers are weak due to the limited spatial density of the RE-RE zig-zag-chain-shaped chemical bonds shown in Fig. 1(c). This kind of weak bond interactions can also be found in $KV_3Sb_5$ family of materials. Thus, similar to the cleavable lattice layers in $KV_3Sb_5$ family, it is expected that $EuTi_3Bi_4$ can also be exfoliated by knives or tapes, and thin flakes of $EuTi_3Bi_4$ can be obtained to make device fabricating available. The crystals of $EuTi_3Bi_4$, $GdTi_3Bi_4$, and $SmTi_3Bi_4$ are grown by flux method as seen in Fig. 1(f). Fig. 1(d) shows the single crystal X-ray diffraction (XRD) pattern of the hexagonal surface of $RETi_3Bi_4$, indicating the high quality of $RETi_3Bi_4$ with $c \approx$ 25.46 Å, 24.81 Å and 24.89 Å for Eu, Gd and Sm. Although these three samples are adjoined each other on periodic table, the length of c axis of $EuTi_3Bi_4$ are significantly larger than other two samples, which might be caused by the abnormal lanthanide contraction of Eu. Fig. 1(e) displays the energy dispersive spectra (EDS) for samples with different rare earth elements as Eu, Gd and Sm. The ratio of RE: Ti: Bi is nearly 1:3:4, indicating a stoichiometric atomic ratio.

Due to the weak interlayer interactions between two adjacent RE/Bi layers, $RETi_3Bi_4$ family is expected to be cleavable by knives or tapes. To verify it, by repeatedly stacking on the crystal surface with two pieces of tape (Nitto, BT-150E-KL), large number of shiny pieces are symmetrically distributed on both sides of the tapes (Fig. 2(a)), indicating the sufficient exfoliation of $RETi_3Bi_4$. By pressing the tape onto silicon wafers with 300 nm $SiO_2$ on top, one can further obtain $RETi_3Bi_4$ nano-flakes with dozens of micrometers on lateral sizes as displayed in Fig.2(b)-(d). The flakes all show blue or cyan color, indicating the thin thickness as low as nanometers. By scanning the thickness of these flakes through an atomic force microscope (AFM), one can see that the thickness are varies from 5 nm to dozens of nanometers, as displayed by the red line-scanning curves overlay on the respective optical images in Fig. 2(b)-(d). These results confirm that $RETi_3Bi_4$ family is cleavable and suggest the feasibility for mesoscopic device fabrications.

We then study the magnetic properties of $RETi_3Bi_4$ family utilizing a vibrating sample magnetometer (VSM). The results are summarized in Fig. 3. Fig. 3(a)-(c) display the magnetization versus external magnetic field (*M-H*) of $EuTi_3Bi_4$ (H//c), $GdTi_3Bi_4$ (H//c), and $SmTi_3Bi_4$ (H//ab plane) respectively under various temperatures from 2 K to 50 K. In Fig. 3(a), one can clearly see a typical soft out-of-plane ferromagnetic type *M-H* curves of $EuTi_3Bi_4$. A hysteresis loop with coercive field 130 Oe is received under the temperature of 2 K, and the magnetic moments are saturated when increasing the magnetic field up to 0.4 T. The saturation magnetization ($M_s$) of EuTi$_3$Bi$_4$ is ~ 5.3 $\mu_B$/f. u. taking into the six 4f electrons of Eu$^{3+}$, and the results are consistent with a ferromagnetic order of EuTi$_3$Bi$_4$. The $M_s$ provided by Eu about 5$\mu_B$ are also be observed in recent research of the EuV$_3$Sb$_4$, the variation of RETi$_3$Bi$_4$[41]. The inset displays magnetic moments saturates under lower fields when applying fields along c axis (*H*//c) comparing with the case for *H*//ab plane, pointing out that for EuTi$_3$Bi$_4$, the easy axis are along c axis. Fig. 3(b) shows the magnetization versus temperature under a fixed magnetic field of 100 Oe. By sweeping the temperature through a zero-field-cooling (ZFC) and a field-cooling (FC) processes, one can see that the magnetization rises quickly when the temperatures are lower than 15 K, and a typical divarication between ZFC and FC curves of ferromagnetic order can be identified. By taking the 1st derivative of the FC curve, Curie temperature ($T_C$) of 11 K is obtained by extracting the temperature value at the extremum of the derivative curve. For the *M-H* curves of GdTi$_3$Bi$_4$ in Fig. 3(c), one can identify that GdTi$_3$Bi$_4$ holds an out-of-plane AFM order with zero net magnetization under zero magnetic field. By applying magnetic fields, two steps of spin flop transitions with the critical fields of $H_{c1}$ = 1.6 T and $H_{c2}$ = 3.3 T can be observed. This kind of multiple spin flop transitions are also previous reported in several AFM magnets[27,42-44]. Saturation magnetization $M_s$ of ~ 6.7 $\mu_B$/f. u., which is also close to the ideal 7 $\mu_B$ from Gd$^{3+}$ ions, is received when the field is larger than 6 T in GdTi$_3$Bi$_4$. The inset of Fig. 3(c) displays the magnetic moments saturates under lower fields when applying fields along c axis (*H*//c) and for *H*//ab plane, indicating that for GdTi$_3$Bi$_4$, the easy axis are along c axis, same as in EuTi$_3$Bi$_4$. The FC and ZFC curves of GdTi$_3$Bi$_4$ in Fig. 3(d) coincide exactly under all temperatures, and a representative kink feature can be identified as marked by the arrow, which indicates the Néel temperature of 14.5 K. Contrary to the out-of-plane magnetic order in EuTi$_3$Bi$_4$ and GdTi$_3$Bi$_4$, SmTi$_3$Bi$_4$ shows rather small in-plane ferromagnetism under low temperatures and high magnetic field in Fig. 3(e). The inset displays the magnetic moments saturates under lower fields when applying fields along c axis (*H*//c) comparing with the case for *H*//ab plane, and the orbital magnetic moment may cause the diamagnetic contribution of the negative signal when *H*//ab. The magnetization versus temperature of SmTi$_3$Bi$_4$ under a fixed magnetic field of 100 Oe are displayed in Fig. 3(f). The ZFC and FC curves show a ferromagnetic order, and the 1st derivative of the FC curve, Curie temperature ($T_C$) of 23.3 K is also obtained.

One may notice that the magnetism in RETi$_3$Bi$_4$ family is simple comparing with those magnetic kagome materials with complex magnetic orders, and the $M_s$ value of the samples with spontaneous magnetization are relatively large for EuTi$_3$Bi$_4$ and GdTi$_3$Bi$_4$. We attribute these two points to the reason that only rare earth elements in RETi$_3$Bi$_4$ family contribute the magnetic order. The magnetic moments containing in each rare earth atom or kation are relatively larger than the ones in transition metal elements. In addition, the factor that Ti kagome lattices in RETi$_3$Bi$_4$ do not contribute the magnetic order also benefits the formation of a simple magnetic order with large $M_s$. The reason is that the moments provided by the kagome lattices formed by magnetic elements may align oppositely to the ones provided by rare earth elements[19], forming frustrated magnetism[11], or forming complex spiral magnetisms[45].

The band structures of RETi$_3$Bi$_4$ family are also investigated in this work. Here, we take EuTi$_3$Bi$_4$ as the representative to elucidate the exotic electronic properties of these materials with kagome sublattices. Fig. 4(a) presents the first Brillouin zone, the high-symmetry k points, and the corresponding two-dimensional projected surface of EuTi$_3$Bi$_4$ primitive cell. The calculated results are in great agreement with the experimental results when U = 6 eV by comparing. Fig. 4(b) illustrates the band structure of EuTi$_3$Bi$_4$ with SOC in FM state. Obviously, we can observe the bands with van Hove singularity (vHs)-like at point Y known as saddle-type bands. For the sake of greater clarity, we calculate the bands contributed by 3$d$ orbitals of Ti atoms forming the kagome sublattice along a high-symmetric $k$ path as close to the hexagon Brillouin zone (BZ) as possible, as depicted in Fig. 4(c). The typical band characteristics caused by the kagome structure are roughly plotted with blue lines in Fig. 4(c). Around the point C$_0$, we discover the distinct bands with Dirac cone-like above the Fermi level about 0.1 eV and the flat bands are located above them about 1.5 eV. These calculation results are typical characteristic of bands for the materials with kagome sublattice.

For further exploring the surface electronic properties of EuTi$_3$Bi$_4$, we calculate the surface states from bulk part and the corresponding Fermi surface on 2D projected surface of EuTi$_3$Bi$_4$ as illustrated in Fig. 4(d) and (e), respectively. A similar kagome bands can also be discovered in Fig. 4(d). In Fig. 4(e), we can observe six petal-like

Fermi surface patterns. These results are agreed well with the experimental measurements as follows. However, due to the predicted Dirac cone-like bands are above the Fermi level, it is difficult to be observed in experimental measurements.

We then investigate the band structure of RETi$_3$Bi$_4$ via ARPES measurement. By mapping out the ARPES dispersions detected under various photon energies, the out-of-plane ($k_x$-$k_z$ plane) constant energy contours (CECs) near E$_F$ can be obtained as seen in Fig. 5(a). The CECs along $k_z$ direction show no clear cyclical change along with BZ, indicating the quasi-2D property of the dispersion in RETi$_3$Bi$_4$. In Fig. 5(b), the Fermi surface structures of EuTi$_3$Bi$_4$, GdTi$_3$Bi$_4$, and SmTi$_3$Bi$_4$ are demonstrated. One can clearly see similar hexagonal contours in all three materials. For EuTi$_3$Bi$_4$, one can see the strong signals at the K points of the first BZ which correspond to the six petal-like Fermi surface patterns obtained by calculations in Fig. 4(e). The ring-shaped pocket surrounding Γ point and the finite density of states at Γ point can also be identified in the second BZ. These results are consistent with the calculation demonstrated above. The finite density of states at Γ point in EuTi$_3$Bi$_4$ evolves to an annular-shaped pocket in the case of GdTi$_3$Bi$_4$ (red dotted circle) and SmTi$_3$Bi$_4$, indicating the Fermi energy of GdTi$_3$Bi$_4$, and SmTi$_3$Bi$_4$ is higher than EuTi$_3$Bi$_4$ since an electron pocket is expected at Γ point according to the calculation. Fig. 5(c)-(e) displays the detected band dispersions along Γ- C$_0$ direction of EuTi$_3$Bi$_4$, GdTi$_3$Bi$_4$, and SmTi$_3$Bi$_4$ respectively. By overlay the calculation on the ARPES results in the Eu case in Fig. 5(c), we can find that the ARPES results are quantitatively match with the calculated dispersions with a large V-shaped dispersion crossing C$_0$-Γ-C$_0$, indicating the veracity of the band calculations in Fig. 4. Additionally, the flat bands caused by Eu are also observed between -1.5 and -2.0 eV in Fig. 5(c). In Fig. 5(d) and (e), band dispersions of GdTi$_3$Bi$_4$ and SmTi$_3$Bi$_4$ are similar to those in EuTi$_3$Bi$_4$. The difference is that the Fermi energy is ~ 0.3 eV higher than EuTi$_3$Bi$_4$, and the conduction band minimum of a branch of conduction band at Γ point is lower than E$_F$ in GdTi$_3$Bi$_4$, becoming detectable by our ARPES measurement as marked by the red dotted curves in Fig.5(d).

**Summary**

In conclusion, we systematically investigate a new family RETi$_3$Bi$_4$ (RE = Eu, Gd and Sm) with weak interlayer interactions and clean kagome layers via magnetic measurement, ARPES measurement and first-principles calculation. The weak interlayer interactions between adjacent RE/Bi caused by the limited spatial density of the RE-RE zig-zag chain shaped chemical bonds make it easy to do mechanical exfoliation until nanometer scale. The rare earth elements solely provide the magnetic moments in RETi$_3$Bi$_4$, and magnetisms of FM, AFM, and PM are formed for RE = Eu, Gd, and Sm respectively. The magnetic orders are simple and the saturation magnetization are relatively large in EuTi$_3$Bi$_4$ and GdTi$_3$Bi$_4$ comparing with the previous studied transition metal based kagome lattices. The first-principle calculations and ARPES are consistent with each other. We provide insights into the electronic structure of RETi$_3$Bi$_4$ family, suggesting an ideal platform to investigate the exotic magnetic topological phases and the related topological quantum transport studies. We believe that this new Kagome magnets family is worthy of further studies on kagome physics and offer a wide range of potential applications in the future.

*Add note*: We noticed a recent study on the same family of material RETi$_3$Bi$_4$ (RE = Yb, Pr, and Nd)[46], in which the tunable magnetism and electron correlation are studied. These two studies complement each other and show more plentiful properties in this family of material.

**Methods**

**Crystal Growth.** The flux method is utilized to synthesize the high-quality single crystals. The starting elements of Eu (Gd, Sm): Ti: Bi in the molar ratio of 1.2: 1: 20 are mixed in an alumina crucible and sealed in a quartz ampule. The ampule is placed in a furnace and heated to 1000 °C at a rate of 1 °C/min. After maintaining it at 1000 °C for 16 h, the ampule is slowly cooled down to 520 °C in 2000 min. The hexagonal shaped single crystals are obtained after centrifuging in order to remove the excess flux.

**Magnetic Measurements.** Magnetic properties are measured with a Physical Properties Measurement System (Quantum Design) in the temperature range 2 K ≤ $T$

≤ 300 K with magnetic field up to 9 T. Standard copper holders and quartz holders are used during the measurements when $H//c$ and $H//ab$ respectively.

**ARPES.** ARPES experiments are performed at the BL03U and the BL09U beamlines at the Shanghai Synchrotron Radiation Facility. Data is collected with photon energies ranging from 48 to 120 eV with linear horizontal polarization. The samples are cleaved and measured in an ultrahigh-vacuum chamber at a pressure below $10^{-10}$ Torr.

**Band Structures Calculations.** The band structures calculations of $EuTi_3Bi_4$ are performed by the Vienna ab initio Simulation package (VASP) with the projector augmented-wave (PAW) formalism based on the density functional theory (DFT) [47,48]. The Perdew-Burke-Ernzerhof (PBE) exchange-correlation functional with the generalized gradient approximation corrected for on-site Coulombic interactions (GGA + $U$) method is adopted [49,50]. The parameter $U$ is selected to be 6 eV to treat the effect of Coulomb correction from the localized 4$f$ orbitals of Eu atoms. The cutoff energy for the plane-wave basis is set at 520 eV and the Brillouin zone (BZ) integral is sampled by 7 × 7 × 7 Γ-centered $k$ mesh. The local magnetic moments on Eu atoms for ferromagnetic state (FM) are along the $z$ axis. Spin-orbit coupling (SOC) is taken into account in all calculations. The tight-binding model of $EuTi_3Bi_4$ is constructed by the Wannier90 with Eu 4$f$ orbitals, Ti 3$d$ orbitals, and Bi 6$p$ orbitals based on the maximally-localized Wannier functions (MLWF) [51]. The surface states and the corresponding Fermi surface are calculated by using the WannierTools software package [52].

**Acknowledgments**

The authors gratefully acknowledge the financial support of the National Key Research and Development Program of China (Grant No. 2022YFA1402404); the National Natural Science Foundation of China (Grant Nos. 92161201, T2221003, 12104221, 12104220, 12274208, 12025404, 12004174, 91961101, 61822403, 11874203, 12104220); the Fundamental Research Funds for the Central Universities

(Grant No. 020414380192). Part of this research used Beamline 03U of the Shanghai Synchrotron Radiation Facility, which is supported by ME2 project under Contract No.11227902 from National Natural Science Foundation of China. The authors also thank Professor Peng Zhang from Nanjing University for discussions.

**Data availability**

The datasets that support the findings of this study are available from the corresponding author upon reasonable request.

**Competing interests**
The authors declare no competing financial interests.

# Figures

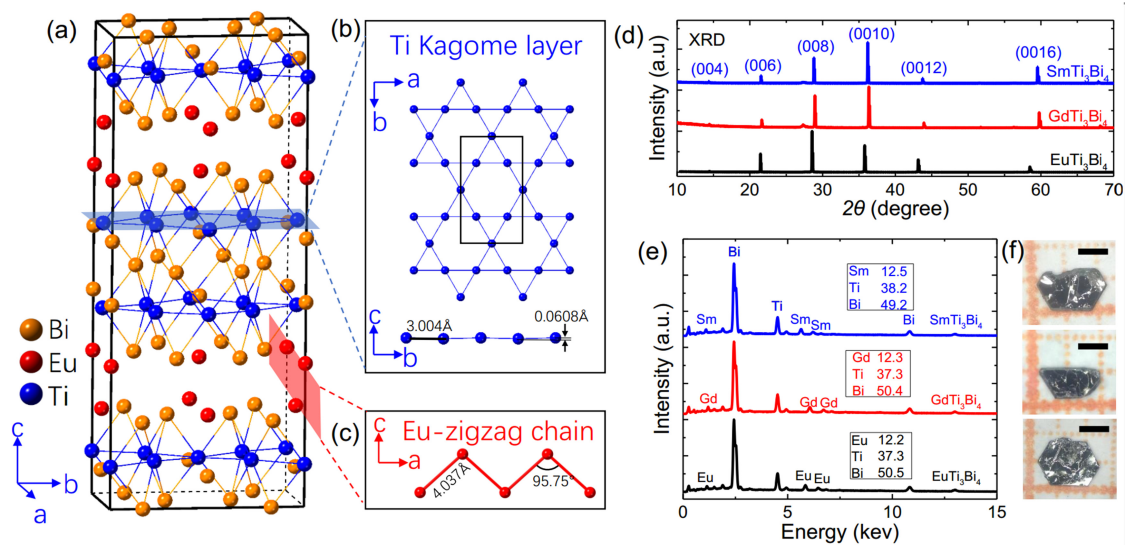

**Fig. 1. Crystal structures and characterizations.** (a) Crystal structure diagrams of EuTi$_3$Bi$_4$. (b)The diagram of Ti kagome layer with slightly distortion along c axis. (c) The diagram of zigzag chain formed by rare earth elements (d) Single crystal XRD pattern for RETi$_3$Bi$_4$ samples. (e) EDS results of RETi$_3$Bi$_4$. (f) Optical photographs of as-grown RETi$_3$Bi$_4$ crystals. Scale bar: 1 mm.

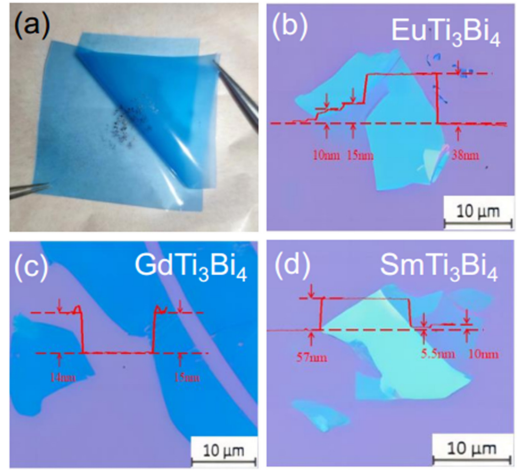

**Fig. 2. Mechanical exfoliation of RETi$_3$Bi$_4$.** (a) The demonstration of mechanical exfoliation of RETi$_3$Bi$_4$ by tapes. (b-d) Optical images and the corresponding AFM line-scanning profiles of the few-layer flakes of EuTi$_3$Bi$_4$, GdTi$_3$Bi$_4$, and SmTi$_3$Bi$_4$ respectively.

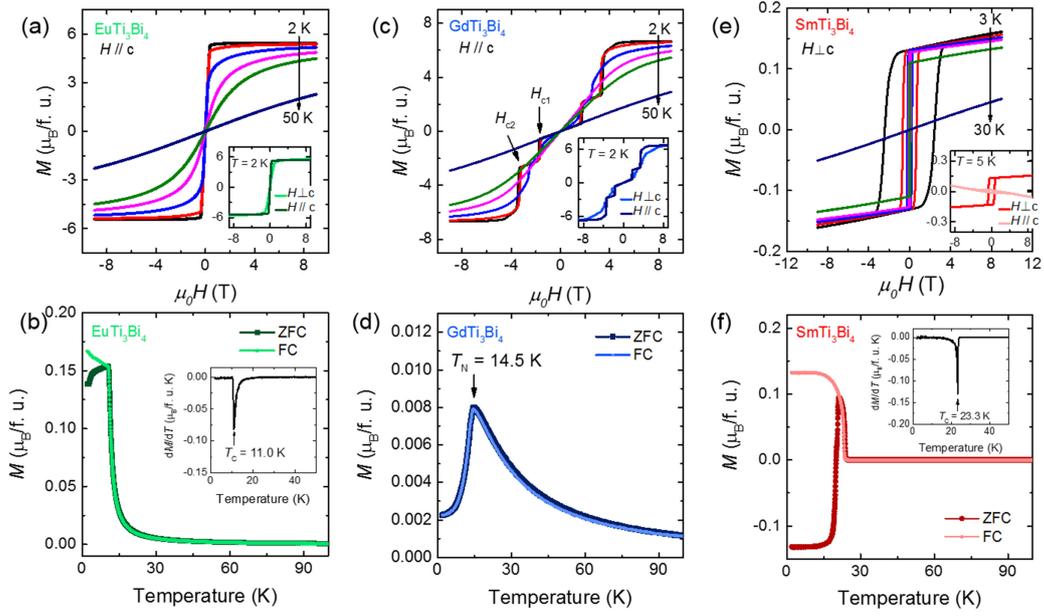

**Fig. 3. Magnetic measurements.** (a) Magnetization versus magnetic field for EuTi$_3$Bi$_4$ at 2 K, 5 K, 10 K, 15 K, 20 K and 50 K, $H//c$. The inset displays *M-H* curves along c axis ($H//c$) and $H//ab$ plane of EuTi$_3$Bi$_4$ at 2 K. (b) FC-ZFC curves of EuTi$_3$Bi$_4$ under an external magnetic field of 100 Oe ($H//c$). The inset displays the 1st derivative of FC curves. (c) The same as (a) but for GdTi$_3$Bi$_4$. (d) The same as (b) but for GdTi$_3$Bi$_4$. (e) *M-H* curves for SmTi$_3$Bi$_4$ at 3 K, 5 K, 10 K, 15 K, 20 K and 30 K, $H//ab$. The inset displays the *M-H* curves along c axis ($H//c$) and $H//ab$ plane of SmTi$_3$Bi$_4$ at 5 K. (f) FC-ZFC curves of SmTi$_3$Bi$_4$ under an external magnetic field of 100 Oe ($H//ab$). The inset displays the 1st derivative of FC curves.

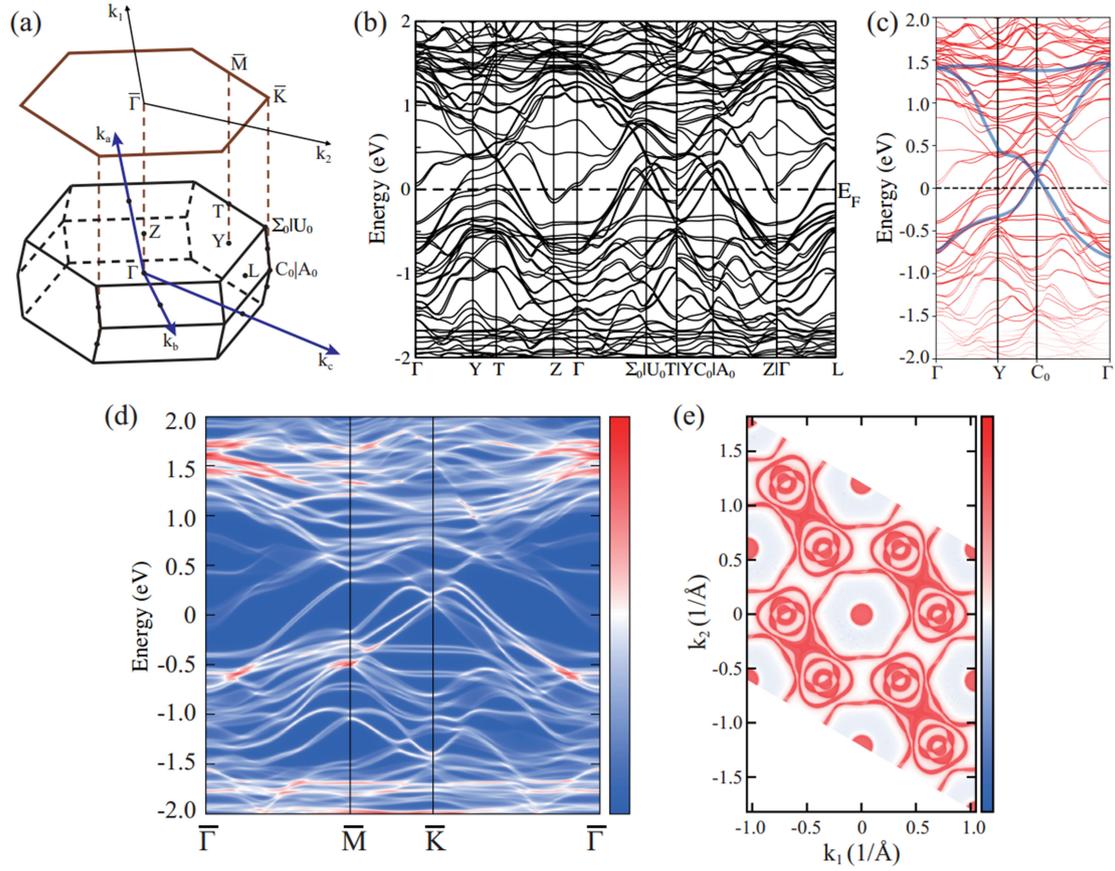

**Fig. 4. Electronic structures calculations.** (a) The first Brillouin zone (BZ) and two-dimensional projected surface of EuTi$_3$Bi$_4$. (b) The bulk band structure of EuTi$_3$Bi$_4$ with SOC (U=6 eV). (c) The band projection contributed by $3d$ orbitals of Ti atoms forming the kagome sublattice. The blue lines represent a schematic of the typical bands resulting from the kagome structure. (d-e) The surface states from bulk part (d) and the corresponding Fermi surface (e) on the two-dimensional projected surface of EuTi$_3$Bi$_4$ with SOC (U = 6 eV).

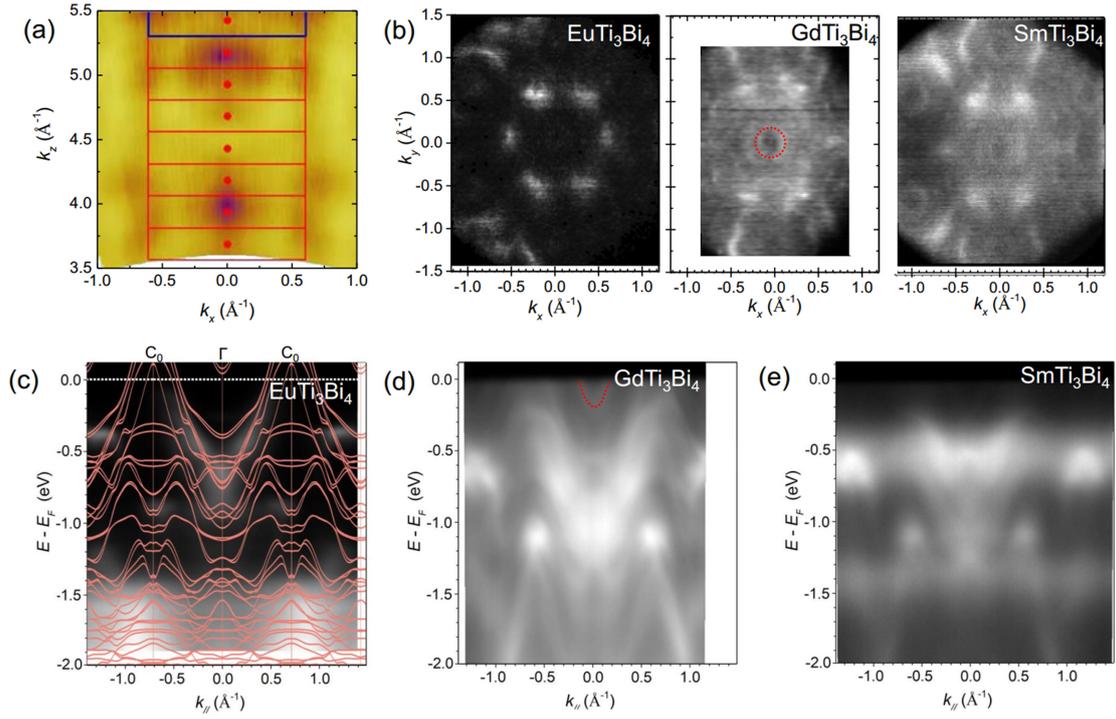

**Fig. 5. ARPES measurements.** (a) constant energy contours near Fermi energy of SmTi$_3$Bi$_4$ in $k_z - k_x$ plane. (b) Fermi pocket morphology of EuTi$_3$Bi$_4$ (left), GdTi$_3$Bi$_4$ (middle), and SmTi$_3$Bi$_4$ (right), (c-e) band dispersion along C$_0$-Γ-C$_0$ direction of (e) EuTi$_3$Bi$_4$ (f) GdTi$_3$Bi$_4$ and (g) SmTi$_3$Bi$_4$ measured with photon energy of 100 eV. All of the spectra are collected at the temperature of 20 K.